\documentclass{article}

\usepackage[a5paper, margin=0.5in]{geometry}

\usepackage[utf8]{inputenc}
\usepackage[none]{hyphenat}
\usepackage{amsmath}
\usepackage{amssymb}
\usepackage{amsfonts}
\usepackage{graphicx}
\usepackage{nicefrac}
\usepackage{mathtools}
\usepackage{verbatim}
\usepackage{multirow}
\usepackage{hhline}
\usepackage{makecell}
\newcommand{\ind}{\perp\!\!\!\!\perp}

\newcommand{\cond}{\,|\,}
\usepackage{setspace}

\newcommand{\expect}{\text{$\mathbb{E}$}}
\newcommand{\E}[2][~]{\ifthenelse{\equal{#1}{~}}{\expect\hspace{-1pt}\left[#2\right]}{\expect_{#1}\left[#2\right]}}
\newcommand{\Eb}[3][~]{\ifthenelse{\equal{#1}{~}}%
{\expect\hspace{-1pt}\left[\hbox{${\left.#2\,\vphantom{#3}\right\cond }$}\hbox{${\left.\,#3\right.}$}\right]}%
{\expect_{#1}\hspace{-1pt}\left[\hbox{${\left.#2\,\vphantom{#3}\right\cond }$}\hbox{${\left.\,#3\right.}$}\right]}}

\newcommand{\prob}{$p$}
\newcommand{\pr}[2][~]{\ifthenelse{\equal{#1}{~}}{\text{\prob}\hspace{-1pt}\left\{#2\right\}}{\equal{#1}{~}}{\text{\prob}_{#1}\left\{#2\right\}}}
\newcommand{\prb}[3][~]{\ifthenelse{\equal{#1}{~}}%
{\mbox{\prob}\hspace{-1pt}\left(\hbox{${\left.#2\,\vphantom{#3}\right|}$}\hbox{${\left.\,#3\right.}$}\right)}%
{\mbox{\prob}_{#1}\left(\hbox{${\left.#2\,\vphantom{#3}\right|}$}\hbox{${\left.\,#3\right.}$}\right)}}
\newcommand{\Prob}{\text{$\mathbb{P}$}}
\renewcommand{\Pr}[2][~]{\ifthenelse{\equal{#1}{~}}%
{\Prob\hspace{-1pt}\left\{\hbox{${\left.#2\right.}$}\right\}}%
{\Prob_{#1}\left\{\hbox{${\left.#2\right.}$}\right\}}}
\newcommand{\Prb}[3][~]{\ifthenelse{\equal{#1}{~}}%
{\Prob\hspace{-1pt}\left\{\hbox{${\left.#2\,\vphantom{#3}\right|}$}\hbox{${\left.\,#3\right.}$}\right\}}%
{\Prob_{#1}\left\{\hbox{${\left.#2\,\vphantom{#3}\right|}$}\hbox{${\left.\,#3\right.}$}\right\}}}

\title{Causal treatment effect decompositions with time-to-event outcomes under competing events}
\usepackage{authblk}
\author[1,2]{Mikko Valtanen}
\author[2]{Tommi Härkänen}
\author[2,3]{Jenni Lehtisalo}
\author[2,3]{Tiia Ngandu}
\author[3,4,5,6]{Miia Kivipelto}
\author[1,5]{Kari Auranen}

\affil[1]{\footnotesize Department of Mathematics and Statistics, 20014 University of Turku, Finland}
\affil[2]{\footnotesize Public Health, Finnish Institute for Health and Welfare, PO Box 30, FI-00271 Helsinki, Finland}
\affil[3]{\footnotesize Institute of Public Health and Clinical Nutrition, University of Eastern Finland, Kuopio, Finland}
\affil[4]{\footnotesize Division of Clinical Geriatrics, Center for Alzheimer Research, Care Sciences and Society (NVS), Karolinska Institutet, Stockholm, Sweden}
\affil[5]{\footnotesize Ageing Epidemiology Research Unit (AGE), School of Public Health, Imperial College London, London, UK}
\affil[6]{\footnotesize Theme Inflammation and Aging, Karolinska University Hospital, Stockholm, Sweden}
\affil[7]{\footnotesize Department of Clinical Medicine, 20014 University of Turku, Finland}
\begin{document}
\date{}
\maketitle

\sloppy
\begin{abstract}
Inference about treatment effects for time-to-event outcomes is often obscured by the presence of competing events. A particularly complex situation arises when the treatment influences the occurrence of the competing event.  
A comprehensive assessment should then account for different mechanisms by which the treatment and the competing event together produce the apparent treatment effect. 
Here, we propose a decomposition of the treatment's effect on the event of interest (target), characterising how it arises due to four distinct mechanisms involving both the target and competing events.
Based on a causal model, the decomposition relies on cross-world estimands reflecting counterfactual scenarios in which the treatment affects the two events as if set to conflicting levels.
We specify exchangeability and consistency assumptions under which the decomposition can be estimated from observed data. We discuss how the new decomposition reveals the role of the competing event and serves as a basis for defining causally interpretable estimands in the presence of competing events.
Finally, we demonstrate the use of the four-way decomposition with datasets from two randomised trials.
\end{abstract}

\section{Introduction}

Analyses of time-to-event outcomes in health research are often complicated by competing events that prevent the event of interest (i.e. target event) from occurring, thereby obscuring interpretation of apparent treatment effects \cite{Prentice1978,Kalbfleisch2002,Andersen2012}. 
Since contrasts of hazards lack a coherent causal interpretation, risk-based effect measures are generally preferable in causal inference settings \cite{Hernan2010,Aalen2015,Sjoelander2016,Martinussen2020,Martinussen2022}. 
Two common risk-based measures include the net risk (i.e. marginal cumulative incidence), which treats the competing events as uninformative censorings, and the cause-specific cumulative risk \cite{Geskus2015}.

A particular problem arises if the treatment also influences the competing event \cite{Geskus2024}. 
The treatment may appear protective against the event of interest simply by accelerating the competing event, thereby leaving less time for the former to occur. 
Conversely, if the treatment delays the competing events, it appears to increase the risk of target events.
Thus, similar total effects may reflect different underlying mechanisms, which may point to different conclusions about the treatment effect.
To characterise treatment effects comprehensively, it is therefore important to distinguish between different underlying mechanisms.

Young et al. \cite{Young2020} framed the cause-specific and net risks as two causally interpretable estimands.
Under certain exchangeability assumptions, the cause-specific risk difference was shown to estimate the average total effect (TE) while the net risk difference was interpreted as a controlled direct effect (CDE) reflecting the treatment's effect under a hypothetical scenario where competing events were eliminated.
CDE thus allows a two-part decomposition of the total effect, where the difference $\text{TE}-\text{CDE}$ represents mechanisms in which the competing event intercepts the occurrence of target events.

Stensrud et al. \cite{Stensrud2020} proposed another two-way decomposition in the competing-risks setting.
Their separable effects framework views the treatment as comprising two components: one affecting only the target event and the other affecting only the competing event.
Separable direct and indirect effects are then defined as contrasts under modified treatments changing only one of the components at a time.
A key strength of this framework is its testable real-world implications.
Nonetheless, requiring a feasible treatment decomposition may be overly restrictive. 
For example, holistic treatments, such as lifestyle interventions, inevitably include aspects that influence the occurrence of both events.

Another way to decompose the total effect includes path-specific effects, which assess the treatment's effect through distinct pathways in a causal diagram \cite{Avin2005}.
Domingo-Relloso investigated certain path-specific effects involving competing risks under an additive hazards model \cite{DomingoRelloso2026}.

Decomposing the treatment effect into its constituent parts is central to mediation analysis, in which a third variable, the mediator, transmits part of the treatment's effect on the outcome \cite{VW2015book}.
In the mediation context, a widely used approach to decompose the total effect relies on cross-world estimands, i.e. potential outcomes under hypothetical interventions assigning the treatment to one level and the mediator to the level it would take under an alternative treatment level.
The total treatment effect is partitioned into the direct part not operating through the mediator and the indirect part that is transmitted through the mediator \cite{Pearl2001,Pearl2014,VanderWeele2016}.
VanderWeele \cite{VanderWeele2014} proposed a further decomposition of the total effect into four parts, separating mechanisms arising from treatment-mediator interactions.

This paper proposes a four-way decomposition of the total treatment effect under competing risks. 
The decomposition comprises the controlled direct effect and three distinct interception effects, characterising how the apparent treatment effect arises due to the interplay of the two events.
Based on a causal model, the decomposition relies on cross-world estimands through a process combining the indicators of the two events under elimination of the other.
The new decomposition reveals the role of the competing event and serves as a basis for defining causally interpretable estimands in the presence of competing events.
These components may be further combined to form different target estimands, including natural direct and indirect effects as special cases.

The paper is organised as follows.
Section 2 presents the data structure and notations.
Section 3 introduces an empirical four-way decomposition of the total effect under competing events. 
Section 4 presents the causal definitions of the four components, based on appropriate cross-world estimands, and provides conditions under which their population-level averages are identifiable from empirical data.
Section 5 illustrates the decomposition using simulation. 
In Section 6 we apply our framework to two real-world trials. 
Section 7 discusses the choice of causal effect estimands. 
Section 8 concludes the paper.

\section{Observations and data structure}

Let $A\in \lbrace a,a^\ast\rbrace$ denote a binary treatment, with $a$ treated and $a^*$ untreated. 
We consider a discrete-time model defined on a sequence of time points, $t_0 < t_1 < \dots < t_K$, at which events and state updates may occur. 
Let $Y=\lbrace Y_k\rbrace_{k=0}^K$ and $D = \lbrace D_k\rbrace_{k=0}^K$ denote the indicator processes associated with two competing events and $C=\lbrace C_k \rbrace^{K}_{k=0}$ the indicator process for censoring.
For each $k$, $Y_k$ and $D_k$ are binary indicators that equal one if the corresponding event has occurred by time $t_k$, and zero otherwise. 
The censoring indicator $C_k$ equals one if an individual is lost to follow-up by time $t_k$ without having experienced either $Y$ or $D$. Let $W=(W_1,\ldots,W_h)$ denote the baseline covariates, and $w$ a joint realisation.

Overbars and underbars denote the histories and futures of the processes, respectively.
For example, $\bar Y_k = (Y_0,\dots,Y_k)$ and $\underline{Y}_k = (Y_k,\dots, Y_K)$.
We take $Y$ to represent the target event whose occurrence terminates the entire follow-up and $D$ its competing event.
We assume the ordering $(C_k$, $D_k$, $Y_k)$ for all $k$. 
All individuals are assumed to be event-free and uncensored at time $t_0$, i.e. $C_0=D_0=Y_0=0$.

\section{Decomposing the total effect}

Let $Y_k^{a'}$ denote the potential outcome of $Y_k$ when, possibly contrary to fact, the treatment is assigned to level $a'\in \{a,a^*\}$. The average total effect (TE) on $Y$ by time $t_k$ is defined as the difference in mean potential outcomes between the treated and untreated levels \cite{Young2020, Rubin1974}: 
\begin{equation}\label{eq_TE_standard}
\E{\hbox{TE}_k} = \E{Y_k^{a} - Y_k^{a^*}} = 
\E{Y_k^{a}} - \E{Y_k^{a^*}}.
\end{equation}
Young et al. \cite{Young2020} showed that under uninformative censoring, the total effect (\ref{eq_TE_standard}) is identifiable from observable quantities as a contrast between two cause-specific cumulative risks. 
We restate their identification formula as
\begin{equation}\label{eq_te_empirical1}
\begin{aligned}
\E{\text{TE}_k} &= \psi^{\text{TE}}_k \\
&=  \sum_w \sum_{s=1}^k \Bigl\{\lambda_{Y;s}(a,w)
\prod_{j = 1}^s \big[ 1-\lambda_{D;j}(a,w)\big]
\big[1-\lambda_{Y;j-1}(a,w)\big]\\
& \qquad \qquad \quad - \lambda_{Y;s}(a^*,w)
\prod_{j = 1}^s \big[ 1-\lambda_{D;j}(a^*,w)\big]
\big[1-\lambda_{Y;j-1}(a^*,w)\big] \Bigr\}P(w)\\
\end{aligned}
\end{equation}
with the cause-specific hazards
$$
\begin{aligned}
\lambda_{Y;s}(a^\prime,w) &= P(Y_s = 1\cond  \bar C_s = D_s=Y_{s-1}=0,A=a^\prime,w), \\
\lambda_{D;s}(a^\prime,w) &= P(D_s = 1\cond \bar C_s = Y_{s-1}=D_{s-1}=0,A=a^\prime,w),
\end{aligned}
$$
and $P(w)$ as the joint distribution of the baseline covariates.
Denoting the net risks \cite{Geskus2015} of events $Y$ and $D$ conditionally on $W=w$ and under treatment $a^\prime\in\lbrace a, a^\ast\rbrace$ by
$$
\begin{aligned}
F_{Y;k}(a',w) &= \sum_{s=1}^k \lambda_{Y;s}(a^\prime,w)
  \prod_{j=1}^{k-1} [1-\lambda_{Y;j}(a^\prime,w)],\\
F_{D;k}(a',w) &=  \sum_{s=1}^k \lambda_{D;s}(a^\prime,w)
  \prod_{j=1}^{k-1} [1-\lambda_{D;j}(a^\prime,w)],
\end{aligned}
$$
we can write \eqref{eq_te_empirical1} as (for details, see Appendix \ref{app_A})
\begin{equation}\label{eq_te_empirical}
\begin{aligned}
\psi^{\text{TE}}_k &= \sum_{w}\sum_{s=1}^k \Bigl\{
[F_{Y;s}(a,w) -F_{Y;s-1}(a,w)][1-F_{D;s}(a,w)]\\
& \qquad \qquad \quad -[F_{Y;s}(a^*,w) -F_{Y;s-1}(a^*,w)][1-F_{D;s}(a^*,w)]\Bigr\} P(w) .
\end{aligned}
\end{equation}

Let $f_{Y;s}(\cdot) = F_{Y;s}(\cdot) - F_{Y;s-1}(\cdot)$ denote the net point probability of the target event occurring at time $t_s$ and $\tilde f(s\cond \cdot,Y_k=1) = f_{Y,s}(\cdot)/F_{Y;k}(\cdot)$, $s=0,\dots,k,$ its conditional distribution given that the event occurs by $t_k$.
The total effect can be further rewritten as
$$
\psi_k^{\text{TE}} = \sum_w \Big\{ F_{Y;k}(a,w) \big[1-E_k(a,a,w) \big]
- F_{Y;k}(a^*,w)\big[ 1-E_k(a^*,a^*,w) \big] \Big\}P(w),
$$
where
$$
\begin{aligned}
E_k(a',a'',w) = \sum_{s=1}^k \tilde f_{Y_k}(s\cond a',w){F_{D;s}(a'',w)} = \E[S \sim \tilde f(s\cond a',w,Y_k=1)]{F_{D;S}(a'', w)}
\end{aligned}
$$
is the probability that the target event, following its conditional net distribution under $A=a'$, would be preceded by the competing event, following its net distribution under $A=a''$.
The two cumulative risks contrasted in the total effect may thus be perceived as the fractions
($1-E_k(\cdot)$) of the net risks of the target event ($F_{Y;k}(\cdot)$) that are not intercepted by the competing event under the given treatment.

Rearranging terms gives

\begin{equation}\label{eq_te_E}
\begin{aligned}
\psi_k^{\text{TE}} & = \sum_w \Big\{ \big[F_{Y;k}(a,w) - F_{Y;k}(a^*,w)\big] -\big[F_{Y;k}(a,w) E_k(a,a,w) \\
& \qquad \qquad -F_{Y;k}(a^*,w)E_k(a^*,a^*,w) \big] \Big\}P(w).
\end{aligned}
\end{equation}
The first bracketed term is the contrast of net risks representing, under certain exchangeability conditions, the controlled direct effect (CDE), i.e. treatment effect in the absence of competing events \cite{Young2020}. 
The second bracketed term, TE$-$CDE, captures the part of the total effect involving interception by the competing event. By adding and subtracting terms, this term can be partitioned into three components, leading to a four-way decomposition of the total effect:
\begin{equation*}
\psi^{\text{TE}}_k = 
\psi^{\text{CDE}}_{k} + 
\psi^{\text{INT}_\text{ref}}_{k} + 
\psi^{\text{INT}_\text{med}}_{k} +  
\psi^{\text{PIE}}_k,
\end{equation*}
where 
\begin{equation}\label{eq_decomp_empirical}
\begin{aligned}
\psi^{\text{CDE}}_k &= \quad \sum_{w}\big\{ F_{Y;k}(a,w) - F_{Y;k}(a^*,w)\big\} P(w),\\
\psi^{\text{INT}_\text{ref}}_{k} &= -\sum_{w}\Big\{ F_{Y;k}(a,w)E_{k}(a,a^*,w) - F_{Y;k}(a^*,w)E_{k}(a^*,a^*,w) \Big\} P(w),\\
\psi^{\text{INT}_\text{med}}_{k} &= -\sum_{w}\Big\{F_{Y;k}(a,w)\big[E_{k}(a,a,w) - E_{k}(a,a^*,w \big] \\
& \qquad \qquad - F_{Y;k}(a^*,w)\big[ E_{k}(a^*,a,w) - E_{k}(a^*,a^*,w\big]\Big\} P(w),\\
\psi^{\text{PIE}}_k &= -\sum_{w} \Big\{ F_{Y;k}(a^*,w)\big[ E_{k}(a^*,a,w) - E_{k}(a^*,a^*,w) \big]\Big\} P(w).
\end{aligned}
\end{equation}
Motivated by the analogy to the four-way decomposition in mediation literature \cite{VanderWeele2014}, we refer to these components as the controlled direct effect, reference interception, mediated interception, and the pure indirect effect. 

The four empirical components suggest the following interpretations. 
First, the controlled direct effect $\psi^{\text{CDE}}_k$ expresses the treatment effect in the hypothetical situation where competing events do not occur \cite{Young2020}.
Second, the reference interception $\psi_k^{\text{INT}_\text{ref}}$ 
measures the contribution to the total effect that follows from the interception of the target events by the competing events following their net distribution under the reference treatment $a^*$. 
This effect remains non-zero in the presence of competing events, even when the treatment does not influence those events.
Third, the mediated interception $\psi_k^{\text{INT}_\text{med}}$ captures the change with respect to $\psi_k^{\text{INT}_\text{ref}}$ when the competing event is considered at treatment level $a$ instead of $a^*$. 
This effect is non-zero only if the treatment also affects the competing event.

Finally, the pure indirect effect $\psi_k^{\text{PIE}}$ is the difference in the intercepted portions when the target event follows its net distribution under the reference treatment.
This effect is non-zero when the treatment affects the the competing event, even if it does not affect the target event.

\section{Causal estimands and interpretation}

In this section, we define a causally interpretable four-way decomposition of the total effect and provide 
assumptions under which its population-average components are identifiable from empirical data through expressions \eqref{eq_decomp_empirical}.
The causal definitions are based on cross-world estimands, constructed through a joint process that combine indicators of the two events under elimination of the other.

\subsection{Joint process and cross-world estimands}\label{section_jointprocess}

Let $Y_k^{a',\varnothing}$ denote the target event indicator under a hypothetical intervention where the treatment was set to level $a'$ but the competing event process was somehow eliminated (i.e., prevented from occurring).
Given this setting, Young et al. \cite{Young2020} defined the average controlled direct effect (CDE) up to time $t_k$ as
\begin{equation}\label{eq_cde}
\E{\hbox{CDE}_k} =\E{Y_k^{a,\varnothing} - Y_k^{a^*,\varnothing}} = \E{Y_k^{a,\varnothing}} - \E{Y_k^{a^*,\varnothing}}.
\end{equation}
We term the processes $Y^{a',\varnothing} = 
\lbrace Y_k^{a',\varnothing}\rbrace_{k=0}^K$ and $D^{\varnothing,a'} = \lbrace D_k^{\varnothing,a'}\rbrace_{k=0}^K$ the controlled processes of $Y$ and $D$, respectively, under treatment $a' \in \{a,a^*\}$. 
The corresponding controlled event times are $T^{a'}_Y = \min\lbrace t_k; Y_k^{a',\varnothing}=1\rbrace$ and $T^{a'}_D = \min\lbrace t_k; D_k^{\varnothing,a'}=1\rbrace$.

We define a joint process 
$Z^{a',a''}  = 
\{(D_k^{a',a''},Y_k^{a',a''})\}_{k=0}^K$ as a sequence of indicator pairs from the two controlled processes $D^{\varnothing,a''}$ and $Y^{a',\varnothing}$ until either takes the value $1$, or until time $t_K$ is reached. When either component jumps to one, the joint process remains constant thereafter.
Specifically, if $t_Y^{a'} < t_D^{a''}$, then $D_k^{a',a''} = 0$ and $Y_k^{\alpha_Y,\alpha_D} = Y_k^{\alpha_Y,\varnothing}$ for all $k=0,...,K$. 
If $t_D^{a''}\leq t_Y^{a'}$, then $D_k^{a',a''} = D_k^{\varnothing,a''}$ and $Y_k^{a',\alpha_D} = 0$ for all $k=0,...,K$. Consequently, for any $k$,
\begin{equation}\label{eq_yd}
    Y_k^{a',a''} = Y_k^{a',\varnothing}[1-D_k^{a', a''}].
\end{equation}

The joint process corresponding to treatment levels $(a',a'')$ with $a'\neq a''$  is a counterfactual amalgamation of two controlled processes under the influence of different treatment levels. 
Correspondingly, the joint process with $a' = a''$ combines the $Y$ and $D$ processes under the same treatment level. 
The joint processes $Z^{a',\varnothing}$ and $Z^{\varnothing,a'}$ are defined as the controlled processes $Y^{a',\varnothing}$ and $D^{\varnothing,a'}$.

A joint process with different treatment assignments for the two controlled processes is a cross-world estimand, which may be used to decompose the total effect into parts reflecting its different mechanisms.
In particular, the cross-world quantity corresponding to the target process, it developing under treatment $a'$ and the competing process under treatment $a''$, is $Y_k^{a',a''}$, i.e. the $k$th $Y$ component of the joint process $Z^{a',a''}$.

\subsection{Consistency assumptions and the total effect}

We make the following counterfactual consistency assumptions throughout the paper:
\begin{description}
\item[C1] $D_k^{a'}=0 \implies Y_k^{a'} = Y_k^{a',\varnothing}, \quad k=0,\dots,K$,
\item[C2] $Y_k^{a'}=0 \implies D_{k+1}^{a'}=D_{k+1}^{\varnothing,a'}, \quad k=0,\dots,K-1$.
\end{description}
These assumptions state that if an individual survives one of the two events up to $t_k$ under a given treatment, the subsequent potential outcome of the other event coincides with its corresponding controlled-process indicator.
This implies that the total effect defined in terms of the joint processes agrees with the standard definition \eqref{eq_TE_standard}:
$$
\E{Y_k^{a,a}} - \E{Y_k^{a^*,a^*}}=\E{Y_k^a}-\E{Y_k^{a^*}} =\E{\text{TE}_k} .
$$

For example, as $D_0^{a} = Y_0^{a}=0$ by definition, C2 implies that $D_1^{a}=D_1^{\varnothing,a}$.
If $D_1^{a} = 1$, $Y_k^{a}=0$ and $D_k^{a}=D_k^{\varnothing,a}=1$ for all $k>1$.
If $D_1^{a} = 0$, $Y_1^a=Y_1^{a,\varnothing}$ by C1.
Again, if $Y_1^a = 1$,  $D_k^{a}=0$ and $Y_k^{a}=Y_k^{a,\varnothing}=1$ for all $k>1$.
If $Y_1^a=0$, $D_2^a = D_2^{\varnothing,a}$ by  C2.
Repeating this reasoning up to time $k$, we conclude that $Y_k^a = Y_k^{a,a}$.

\subsection{Causal four-way decomposition of the total effect}

Based on the joint process of the two competing events, the individual-level contrast TE$_k$ ($=Y_k^{a,a} - Y_k^{a^*, a^*}$) of two counterfactual outcomes partitions into the following four-way decomposition:

\begin{equation}\label{eq_decomp}
\begin{aligned}
\text{TE}_k &=  \text{CDE}_k + \text{INT}_{\text{ref},k} + \text{INT}_{\text{med},k} + \text{PIE}_k,
\end{aligned}
\end{equation}
where
\begin{equation}\label{eq_decomposition}
\begin{aligned}
\text{CDE}_k &=                                                Y_k^{a,\varnothing} - Y_k^{a^*,\varnothing}, \\
\text{INT}_{\text{ref},k} &=  (Y_k^{a,a^*} - Y_k^{a^*,a^*}) - (Y_k^{a,\varnothing} - Y_k^{a^*,\varnothing}), \\
\text{INT}_{\text{med},k} &=  (Y_k^{a,a}   - Y_k^{a^*,a})   - (Y_k^{a,a^*}         - Y_k^{a^*,a^*}), \\
\text{PIE}_k &=                Y_k^{a^*a}  - Y_k^{a^*,a^*}.
\end{aligned}
\end{equation}
By equation \eqref{eq_yd}, the four effects can be expressed as (see Appendix \ref{app_decomp}):
\begin{equation}
\begin{aligned}
\text{CDE}_k &= \quad Y_k^{a,\varnothing} - Y_k^{a^*,\varnothing}, \\
\text{INT}_{\text{ref},k} &= -(Y_k^{a,\varnothing}D_k^{a,a^*} - Y_k^{a^*,\varnothing}D_k^{a^*,a^*}), \\
\text{INT}_{\text{med},k} &= 
-(Y_k^{a,\varnothing}D_k^{a,a} - Y_k^{a^*,\varnothing}D_k^{a^*,a}) + (Y_k^{a,\varnothing}D_k^{a,a^*} - 
Y_k^{a^*,\varnothing}D_k^{a^*, a^*}), \\
\text{PIE}_k &= -Y_k^{a^*,\varnothing}(D_k^{a^*,a} -D_k^{a^*,a^*}).
\end{aligned}
\end{equation}

These individual-level causal estimands correspond to the population-level empirical expressions \eqref{eq_decomp_empirical}. 
$\text{INT}_{\text{ref},k}$ captures situations in which the untreated-level competing event ($D^{\varnothing,a^*}$) intercepts the occurrence of one of the two controlled  $Y$ indicators, i.e. cases where one of the counterfactual $Y$ events would have occurred before $t_k$ but is precluded by the competing event.
$\text{INT}_{\text{med},k}$ refers to the change in $\text{INT}_{\text{ref},k}$ when the competing event is considered under the treated level ($D^{\varnothing,a}$).
$\text{INT}_{\text{med},k}$ takes a non-zero value if $D^{\varnothing,a}$ additionally intercepts or undoes the interception of one but not the other $Y$.
The pure indirect effect (PIE) reflects the interception of $Y^{a^*,\varnothing}$ and is non-zero when one but not the other of the untreated-level and treated-level competing events intercepts $Y^{a^*,\varnothing}$.
Note that all four components can only take values in $\{-1,0,+1\}$.

\subsection{Identification of average causal effects}\label{section_identification}

Because the individual-level causal effects are inherently unobservable, we need to consider their population-level averages. 
Here, we provide a set of assumptions that allow the estimation of these average effects from observed data. 

\sloppy
For $a',a''\in\lbrace a,a^*\rbrace$ and 
$k\in\lbrace 0,\ldots,K-1\rbrace$, we make the following exchangeability assumptions:
\begin{description}
\item[A1] $(\bar Y_K^{a',\varnothing},\bar D_K^{\varnothing, a'}) \ind A \cond W $
\item[A2] $Y_{k+1}^{a',\varnothing} \ind D_{k+1}^{\varnothing,a''}\cond Y_k^{a',\varnothing}=0,W$
\item[A3] $D_{k+1}^{\varnothing,a''} \ind Y_{k}^{a',\varnothing}\cond D_k^{\varnothing,a''}=0,W$
\item[S1] $(\underline Y_{k+1}^{a',\varnothing},\underline D_{k+1}^{\varnothing,a'}) \ind C_{k+1}\cond\bar C_k = D_k = Y_k = 0, A=a', W$
\end{description}
Assumption A1 states that, conditionally on the baseline covariates, the two controlled processes are independent of treatment assignment $A$.
The sequential exchangeability assumptions A2 and A3 require that, given the baseline covariates and survival from either process until $t_k$, the next two possible events are conditionally independent.

In addition, we pose the following assumptions for all $k$:
\begin{description}
\item[C3] $A=a' \implies D_k = D_k^{a'} \text{ and } Y_k = Y_k^{a'}$
\item[P1] $P(Y_k=D_k=0\cond\bar C_k = Y_{k-1}=D_{k-1}=0,A=a',w) > 0$
\end{description}
Assumption C3 is the standard consistency assumption, stating that if an individual is observed to receive treatment $A=a'$, the observed indicators equal that individuals potential outcomes under that treatment.
Together with C1--C2, C3 implies that observing an individual with treatment $A=a'$ surviving up to $t_k$ amounts to observing that individual's potential outcomes $(D_1^{\varnothing,a'},Y_1^{a',\varnothing},\dots,D_k^{\varnothing,a'},Y_k^{a',\varnothing})$.
The positivity assumption P1 requires that with each $w\in W$ it is possible to survive up to time $t_K$. 

Under the assumptions A1--A3, S1, C1--C3 and P1, the following results hold (for the proofs, see Appendix \ref{appendix_identification}):
\begin{description}
\item[id1] The average controlled direct effect is identified by (see also \cite{Young2020})
$$
\begin{aligned}
&\E{\text{CDE}_k} = \psi_k^{\text{CDE}}\\
&= \sum_w \sum_{s=1}^k \Bigl\{\lambda_{Y;s}(a,w)
\prod_{j = 1}^s
\big[1-\lambda_{Y;j-1}(a,w)\big]\\
& ~ \hspace{2cm} - \lambda_{Y;s}(a^*,w)
\prod_{j = 1}^s
\big[1-\lambda_{Y;j-1}(a^*,w)\big] \Bigr\}P(w) \\
\end{aligned}
$$
\item[id2] The average reference interception is identified by
$$
\begin{aligned}
&\E{\text{INT}_{\text{ref},k}} = \psi_k^{\text{INT}_\text{ref}} \\
&= -\sum_w \sum_{s=1}^k\Big(\lambda_{Y;s}(a,w) \prod_{j=1}^s\big[1-\lambda_{Y;j-1}(a,w)\big]\Big\{ 1 - \prod_{j=1}^s \big[1 - \lambda_{D;j}(a^*,w) \big] \Big\} \\
& \qquad - \lambda_{Y;s}(a^*,w)\prod_{j=1}^s\big[ 1 - \lambda_{Y;j-1}(a^*,w)\big] \Big\{1 - \prod_{j=1}^s \big[ 1-\lambda_{D;j}(a^*,w) \big] \Big\} \Big)P(w)\\
\end{aligned}
$$
\item[id3] The average mediated interception is identified by
$$
\begin{aligned}
& \E{\text{INT}_{\text{med},k}} = \psi_k^{\text{INT}_\text{med}} \\
&= \sum_w \sum_{s=1}^k\Big(\lambda_{Y;s}(a,w) \prod_{j=1}^s\big[1-\lambda_{Y;j-1}(a,w)\big] \\
& \qquad \qquad \times \Big\{ \prod_{j=1}^s \big[1 - \lambda_{D;j}(a,w) \big] - \prod_{j=1}^s \big[1 - \lambda_{D;j}(a^*,w) \big] \Big\} \\
& \qquad - \lambda_{Y;s}(a^*,w)\prod_{j=1}^s\big[ 1 - \lambda_{Y;j-1}(a^*,w)\big] \\
& \qquad \qquad \times \Big\{\prod_{j=1}^s \big[ 1-\lambda_{D;j}(a,w) \big] - \prod_{j=1}^s \big[ 1-\lambda_{D;j}(a^*,w) \big] \Big\} \Big)P(w)\\
\end{aligned}
$$

\item[id4] The average pure indirect effect is identified by
$$
\begin{aligned}
&\E{\text{PIE}_k} = \psi_k^{\text{PIE}} \\
&= \sum_w \sum_{s=1}^k \lambda_{Y;s}(a^*,w)\prod_{j=1}^s\big[1-\lambda_{Y;j-1}(a^*,w)\big] \\
& \qquad \qquad \times \Big\{ \prod_{j=1}^s\big[1-\lambda_{D;j}(a,w)\big] - \prod_{j=1}^s \big[1-\lambda_{D;j}(a^*,w)\big] \Big\} \\[1ex]
\end{aligned}
$$
\end{description}
It trivially follows that 
$\E{\text{TE}_k}$ is identified as the sum of the four components. Estimates of all effects may be obtained by plugging in estimates of the cause-specific hazards.

The identification of the causal effects was based on the identification of the probabilities $P(Y_k^{a',a''}=1)$ and $P(Y_k^{a',\varnothing}=1)$.
It follows that any causal estimand obtained as a function of these probabilities, such as the restricted mean survival time (RMST; see Appendix \ref{app_rmst}), is also identifiable under the same assumptions.

\section{Illustrations}\label{section_illustrations}

\begin{figure}
    \centering
    \includegraphics[width=1\linewidth]{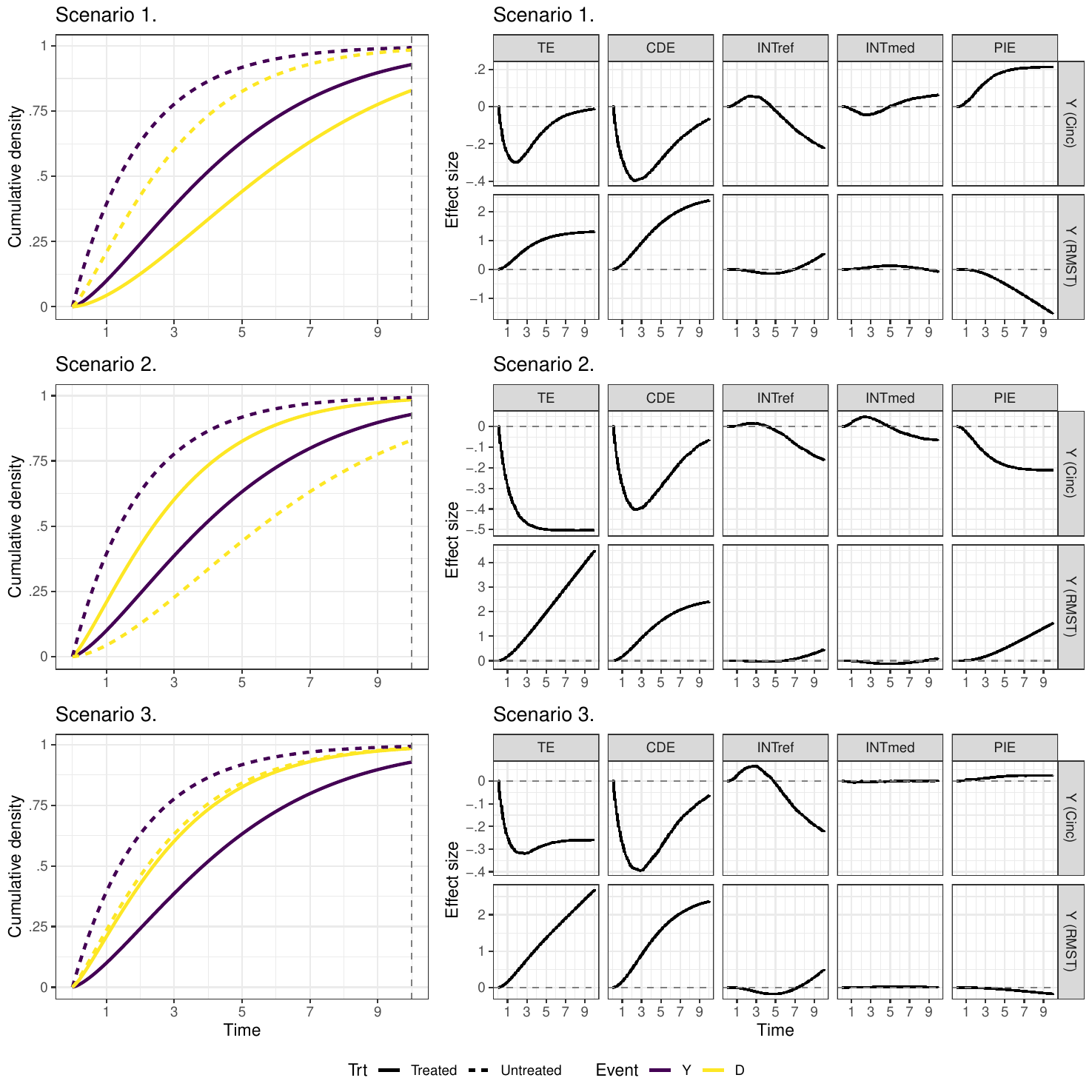}
    \caption{Effect decomposition under three different scenarios. For each scenario, the panels on the left show the net risks of the two events (target $Y$, competing event $D$) under both treatment levels.  
    The panels on the right show the corresponding total effect on $Y$ and its four components over time. The top and bottom rows correspond to the risk and RMST scales, respectively.  In the RMST panels, time refers to the truncation time. The effects were calculated as averages over $n=10,000$ hypothetical individuals, for each of whom the four potential (counterfactual) event times were independently drawn. Negative values on the risk scale indicate beneficial effects, i.e. smaller risks (longer average durations) under treated levels compared to untreated levels.
    }
    \label{fig_illustrate}
\end{figure}

To illustrate the four-way decomposition, we simulated potential outcomes in 10,000 individuals under three scenarios depicting different patterns of interception. 
Figure \ref{fig_illustrate} shows the net risks of the events (left panels) and the population averages of the total effect and its four components (CDE, $\text{INT}_\text{ref}$, $\text{INT}_\text{med}$, PIE) against the target event over time. The effects are presented on the risk (right-panel top rows) and RMST scales (right-panel bottom rows). 

The net risks of the target event ($Y$) are the same across all three scenarios. On the risk scale, 
the controlled direct effect peaks at time $t=3$, after which it gradually wanes towards zero as net risks grow close to one under both treatment levels. On the RMST scale, however, the controlled effect keeps increasing, reflecting the desirable property of RMST in capturing the accumulated effects over the follow-up.

\subsection{Scenario 1: Treatment delays competing event}

This scenario mimics interventions such as lifestyle guidance, which are beneficial against a wide range of adverse health outcomes. 

On average, $\text{INT}_\text{ref}$ is negative, indicating that untreated-level competing event intercepts untreated-level target event more prominently than treated-level target event. Initially, however, $\text{INT}_\text{ref}$ is positive due to the interception of untreated-level target event during early follow-up, when treated-level target event is still rare.

$\text{INT}_\text{med}$ is opposite to $\text{INT}_\text{ref}$,
as the treated-level competing event typically undoes the interception of treated-level target. 
However, as this is not always the case, $\text{INT}_{\text{med}}$ does not fully counteract $\text{INT}_{\text{ref}}$

Although CDE drives the beneficial total effect on the target event within a substantive time window, the total does nevertheless tend to zero due to PIE. Delaying competing events, the treatment gives more chances for target events to occur, thus introducing a harmful contribution on it.

\subsection{Scenario 2: Treatment accelerates competing event}

This scenario corresponds to an intervention in which the treatment aimed against the target event increases the risk of competing event.

$\text{INT}_\text{ref}$ again provides a beneficial contribution to the total. 
The untreated-level target event typically occurs too early for prevention by the untreated-level competing event, whereas the treated-level target event occurs late enough for a substantial portion to be prevented. 
$\text{INT}_\text{med}$ takes the same shape as the reference interception, because the treated-level competing event increases the interception of treated-level target more than the untreated-level target event.

Unlike in Scenario 1, PIE contributes to a beneficial treatment effect.
This arises because the treatment accelerates the onset of the competing event, reducing the available time for the target event to occur.

Eventually, all four components point towards a beneficial treatment effect.
This is duly reflected in the total effect, which quickly starts showing a large protective effect and does not become attenuated over time.
The intuitive explanation is that among the untreated, the target event would almost always occur, while among the treated the occurrence would be much rarer.

\subsection{Scenario 3: Treatment does not affect competing event}

This scenario characterises a treatment designed with a specific effect on the target event with no unwanted side effects.
$\text{INT}_\text{med}$ and PIE are essentially zero, as the treatment does not affect the competing event.
Since the net risk of the untreated-level competing event is the same as in Scenario 1, so is $\text{INT}_\text{ref}$.

This scenario highlights the fact that the competing event plays a role in how the treatment appears to affect the target event even if not itself affected by the treatment.
Even though CDE tends to zero towards the end, a beneficial total effect persists due to $\text{INT}_\text{ref}$.

\section{Empirical data analysis}

To demonstrate the methods in real-world settings, we used data from two randomised trials, in which the treatment either accelerates or delays the competing events. 

\subsection{Treatment delays the competing event}

The Finnish Geriatric Intervention Study to Prevent Cognitive Impairment and Disability (FINGER) \cite{Kivipelto2013} is a randomised clinical trial investigating the effectiveness of a multidomain lifestyle intervention on preventing cognitive impairment.
The study included 1259 participants ($53\%$ male) aged 60--77 years at baseline and deemed at high risk but not yet showing substantial cognitive decline.
We analysed the intervention effect on the pre-specified secondary outcome of cardiovascular disease (CVD), considering other-cause deaths as competing events. A lifestyle intervention can be assumed to lead to holistic health improvement, delaying both the target and competing events.

An earlier study on these data found a protective treatment effect against cardiovascular events among the subpopulation with history of CVD events ($n=145$) \cite{Lehtisalo2022}.
We considered this subpopulation with cardiovascular target events defined as in \cite{Lehtisalo2022}.
Logistic regression was applied to estimate discrete-time hazards using natural cubic splines to allow flexible time-dependence.
The baseline covariates comprised demographic variables (age, sex, education), clinical measurements (systolic blood pressure, total cholesterol, fasting plasma glucose, triglycerides), lifestyle factors (use of blood pressure medication, smoking, BMI) and APOE $\varepsilon 4$ genotype.

\begin{figure}[h]
    \centering
    \includegraphics[width=1\linewidth]{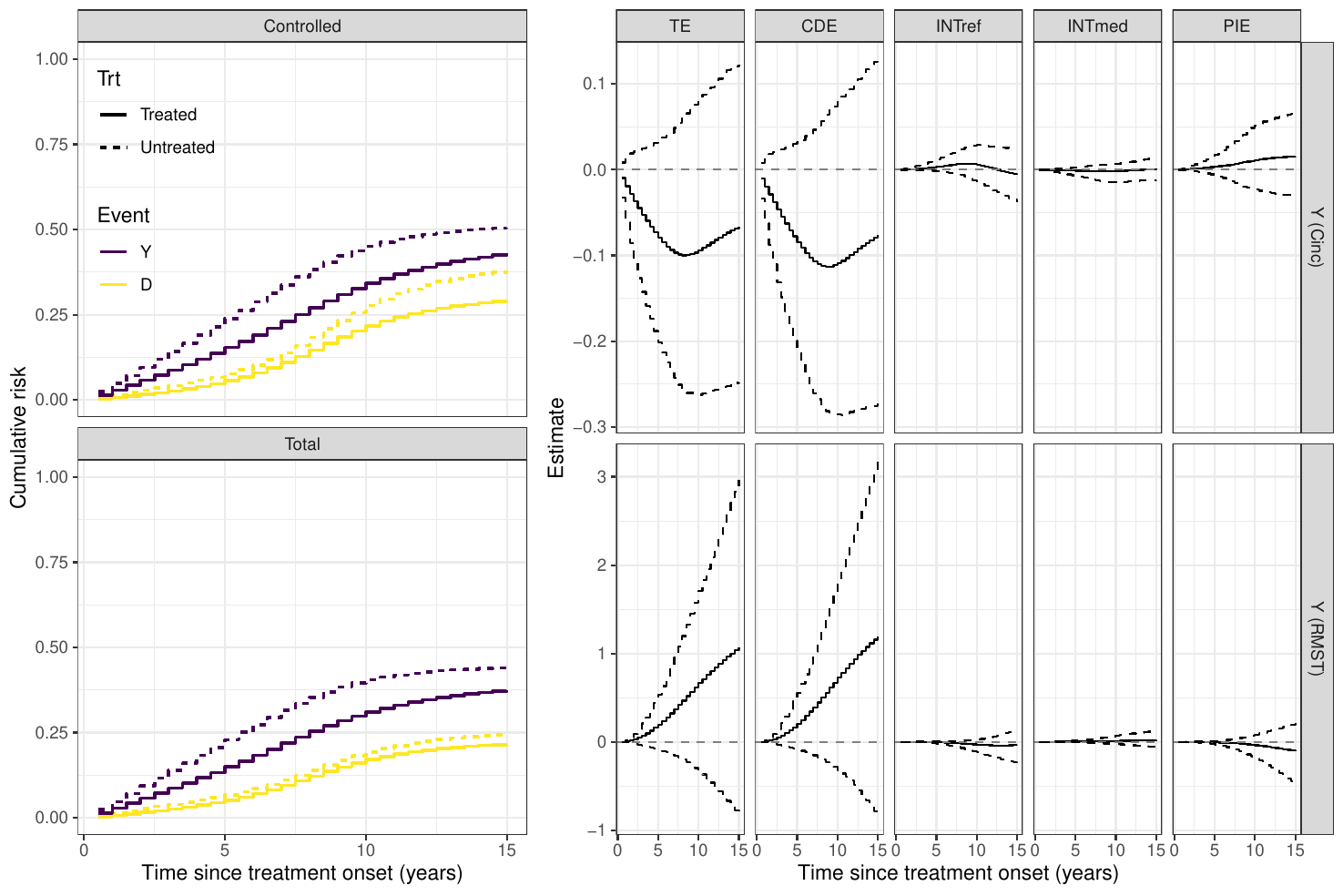}
    \caption{
    Treatment effects on CVD events with other-cause deaths as a competing risk in the FINGER data \cite{Kivipelto2013}.
    Left panel: the estimated net (controlled) and cause-specific (total) risks of cardiovascular events ($Y$) and other-cause deaths ($D$), by treatment group and over time since study onset. Right panel: the estimated decomposition effects for cardiovascular events on the risk and RMST scales (solid lines) and their $95\%$ confidence intervals (dashed lines), over time since study onset. In the RMST panels, time refers to the truncation time.
    }
    \label{fig_finger}
\end{figure}

Figure \ref{fig_finger} shows the estimated decomposition effects for cardiovascular events on the risk and RMST scales over time.
The CDE dominates, as other-cause deaths are rare and interception is consequently minimal.
The total effect on the risk scale peaks at around 7 years of follow-up, reaching a 10 percentage-point decrease in risk.
On the RMST scale, the total effect corresponds to one-year increase in 
cardiovascular-disease-free time over a 15-year time horizon.
A small pure indirect effect arises as the treatment also delays deaths, which allows more time for cardiovascular events to emerge.
The confidence intervals are very wide, but judging by the point estimates, the analysis suggests that other-cause mortality essentially does not affect the treatment effect on CVD events.

\subsection{Treatment accelerates the competing event}
We used publicly available data from a prostate cancer trial \cite{Byar1980} with 502 patients randomised to placebo or to one of three groups receiving varying estrogen (diethylstilbestrol, DES) doses. 
Death from prostate cancer is considered the event of interest, with other deaths as competing events. 
These data were previously used to illustrate causal effect estimands in the presence of competing events with the treatment increasing the risk of competing events \cite{Young2020,Stensrud2020}.

For comparability with the previous analyses, we contrasted the highest-dose DES ($n=125)$ and placebo ($n = 127$) groups. Following Stensrud et al. \cite{Stensrud2020}, we used daily activity function, age group, hemoglobin level, and previous cardiovascular disease as baseline covariates. Logistic regression was applied to model discrete-time hazards using natural cubic splines to allow a flexible time-dependent treatment effect.

\begin{figure}
    \centering
    \includegraphics[width=1\linewidth]{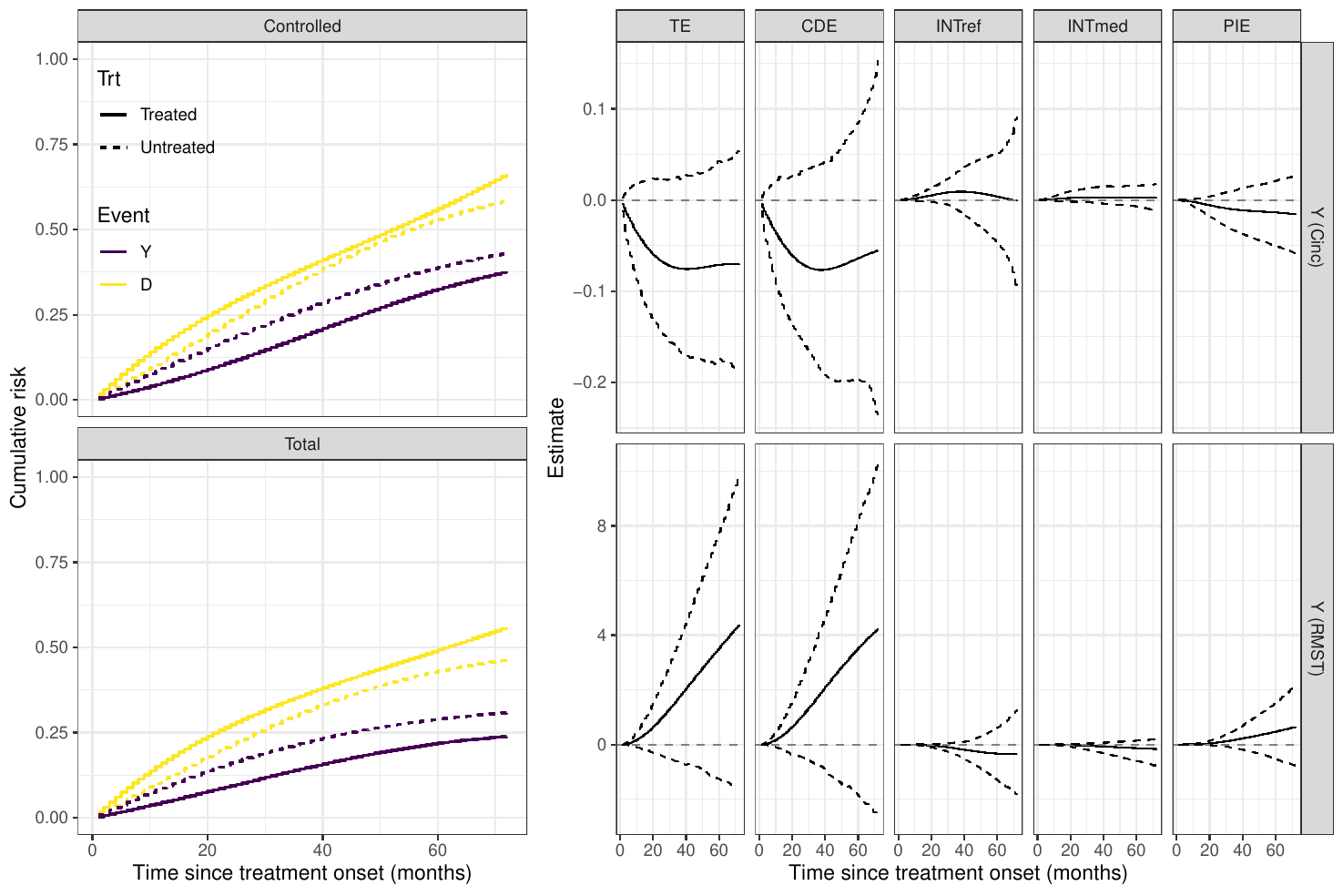}
    \caption{
    Treatment effects on prostate cancer death with other-cause deaths as a competing risk in the prostate data \cite{Byar1980}.
    Left panel: the estimated net (controlled) and cause-specific (total) risks of cardiovascular events ($Y$) and other-cause deaths ($D$), by treatment group and over time since study onset. Right panel: the estimated decomposition effects for cardiovascular events on the risk and RMST scales (solid lines) and their $95\%$ confidence intervals (dashed lines), over time since study onset.
    In the RMST panels, time refers to the truncation time.
    }
    \label{fig_prostate}
\end{figure}

Figure \ref{fig_prostate} 
shows the estimated decomposition effects for prostate cancer on the risk and RMST scales over time.
The CDE on prostate cancer increases during early follow-up, peaking around 36 months with an 8 percentage-point decrease in risk but then attenuates towards 5 percentage-points at the end. 
A small reference interception counteracts the CDE since, regardless of treatment, competing events tend to occur earlier than target events, intercepting these under placebo slightly more often than under the high-dose DES. The pure indirect effect contributes to a beneficial treatment effect because by increasing the incidence of death due to other causes, the DES leaves less time for death due to prostate cancer to occur. The confidence intervals are again wide, but the point estimates suggest that only a very small portion of the perceived beneficial effect on prostate cancer mortality was due to treatment increasing other-cause mortality.

\section{Choosing targeted causal effects}\label{section_choosing}

The four components of decomposition \eqref{eq_decomposition} describe how the apparent (i.e. total) treatment effect is influenced by the interplay of the treatment and the two events.
In practice, the research question may not warrant an overtly detailed partition.
Instead, there may be interest in aspects that can be captured by combining successive components in \eqref{eq_decomposition}.
Importantly, such combinations can only take values in $\{-1,0,+1\}$.
The interpretations of different combinations, as outlined below, are analogous to those in the general mediation setting \cite{VanderWeele2014}.

\subsection{Total effect partitions}

Being void of the influence of the competing event, the controlled direct effect (CDE) can be considered to reflect an isolated relationship between treatment and the event of interest.
CDE is a natural estimand in etiological research, or when comparing effects across populations with different incidences of competing events \cite{Geskus2024}.
The remainder, $\text{TE}-\text{CDE}$, reflects the change in the intercepted portion of events of interest, which summarises how the isolated connection is modified by the existence of the competing event in the real world.

The natural direct effect ($\text{NDE} =\text{CDE}+\text{INT}_{\text{ref}}$) is obtained as the sum of the controlled direct effect and the reference interception. 
It captures the part of the total effect not involving the treatment's effect on the competing event.
The natural indirect effect 
($\text{NIE} =\text{INT}_{\text{med}}+ \text{PIE}$)
is the sum of the mediated interception and pure indirect effect. 
It describes the part of the total effect that is mediated by the treatment affecting the competing event.
When the goal of the treatment is to affect the target event, NDE is a suitable estimand quantifying how much of the resulting total is due to the primarily intended mechanism while NIE quantifies the side effects by affecting the competing event as well.

Including the mediated interception in the direct effect leads to the total direct effect (TDE = $\text{CDE}+\text{INT}_{\text{ref}} + \text{INT}_{\text{med}}$).
TDE includes all mechanisms involving a treatment effect on the event of interest.
If the treatment is designed to affect the competing event, PIE would quantify the intended mechanism and TDE the side effects instead.
For example, preventing nosocomial infection could involve actions aimed at expediting hospital discharge (a competing event for nosocomial hospital-acquired infection).

When the treatment does not specifically target either event but aims at a holistic improvement of health, attributing the mediated interception into either the direct or the indirect part seems unwarranted.
One might then consider assessing a three-way decomposition comprising the natural direct effect, mediated interception and the pure indirect effect.
NDE and PIE quantify treatment's effects via the event-specific mechanisms while $\text{INT}_\text{med}$ is the additional part that does not fit into either category.

\subsection{Relation to separable effects}

Direct and indirect effects have been introduced also in the separable effects framework, where treatment is assumed to comprise two distinct components, one affecting only the target event and the other only the competing event \cite{Stensrud2020}.
The identifying expressions of separable direct and indirect effects coincide with those of NDE and NIE (see Appendix \ref{app_comb}).
This suggests that the natural direct and indirect effects estimated under our framework can be reinterpreted as their separable counterparts, if one is willing to accept the separable treatment assumption.
Interestingly, also $\text{INT}_{\text{med}}$ and PIE can be given interpretations as separable effects defined as similar contrasts  with their separable-effects counterparts defined by setting the two treatment components in the analogous way.
However, CDE and  $\text{INT}_{\text{ref}}$
are not meaningful concepts in the separable effects framework since they explicitly require the hypothetical interventions eliminating competing events.

\section{Concluding remarks}

In this paper, we proposed a new decomposition for the total treatment effect on the event of interest under competing events.
The decomposition comprised four components reflecting different mechanisms by which treatment and the competing event jointly generate the total treatment effect.

The decomposition was based on cross-world estimands where the two processes develop as if the treatment was set to different levels.
To define such estimands, we introduced the concept of a joint process constructed by combining the indicators of the two events under elimination of the other. 
The components of a joint process can be taken to represent cross-world estimands.

Throughout the paper we considered only confounders measured at baseline.
Indeed, time-varying confounders pose a considerable problem for cross-world estimands even if the confounders are exogenous.
The cross-world estimands would then involve an expectation over the longitudinal trajectories among individuals surviving the two events under conflicting treatment assignments.
Such trajectories evidently cannot be estimated from observed data without additional assumptions.

\bibliographystyle{ieeetr}
\bibliography{bib_a2}

\newpage
\appendix

\section{Empirical total effect decomposition}\label{app_A}
Here, we provide  details on the derivation of the expression \eqref{eq_te_E} in the  main text for the total effect.
For simplicity, we drop the baseline covariates $W$. 
The identifiable expression of the total effect (expression \eqref{eq_te_empirical1} in the main text) is
\begin{equation*}
\begin{aligned}
\psi^{\text{TE}}_k
&=  \sum_{s=1}^k \Bigl\{\lambda_{Y;s}(a)
\prod_{j = 1}^s \big[ 1-\lambda_{D;j}(a)\big]
\big[1-\lambda_{Y;j-1}(a)\big]\\
& \qquad \qquad \quad - \lambda_{Y;s}(a^*)
\prod_{j = 1}^s \big[ 1-\lambda_{D;j}(a^*)\big]
\big[1-\lambda_{Y;j-1}(a^*)\big] \Bigr\}.\\
\end{aligned}
\end{equation*}

The net point probability of the event of interest to occur at time $t_s$ is 
$$
f_{Y;s}(\cdot) = F_{Y;s}(\cdot) - F_{Y;s-1}(\cdot) = \lambda_{Y;s}(\cdot)\prod_{j=1}^s[1-\lambda_{Y;j-1}(\cdot)].
$$
The complement of the net risk of the competing event to occur by time $t_s$ is 
$$
1-F_{D;s}(\cdot) = \prod_{j=1}^s[1-\lambda_{D;j}(\cdot)].
$$
Substituting these into the expression of $\psi_k^{TE}$, we obtain
\begin{equation*}\label{eq_app_2}
\begin{aligned}
\psi^{\text{TE}}_k &= \sum_{s=1}^k \Bigl\{
[F_{Y;s}(a) - F_{Y;s-1}(a)][1-F_{D;s}(a)] - \\
& ~ \hspace{2cm} [F_{Y;s}(a^*) - F_{Y;s-1}(a^*)][1-F_{D;s}(a^*)]\Bigr\} \\
&= \sum_{s=1}^k f_{Y;s}(a)[1-F_{D;s}(a)] - \sum_{s=1}^kf_{Y;s}(a^*)[1-F_{D;s}(a^*)] \\
\end{aligned}
\end{equation*}
Then, we multiply and divide the first part by $F_{Y;k}(a)$ and the second by $F_{Y;k}(a^*)$. 
This yields:
\begin{equation*}
\begin{aligned}
\psi_k^{\text{TE}} &= F_{Y;k}(a) \sum_{s=1}^k \cfrac{f_{Y;s}(a)}{F_{Y;k}(a)}[1-F_{D;s}(a)] - F_{Y;k}(a^*) \sum_{s=1}^k \cfrac{f_{Y;s}(a^*)}{F_{Y;k}(a^*)}[1-F_{D;s}(a^*)] \\[1ex]
&= F_{Y;k}(a) [1-E_k(a,a)] - F_{Y;k}(a^*) [1-E_k(a^*,a^*)],
\end{aligned}
\end{equation*}
where $E_k(\cdot,\cdot) = \sum_{s=1}^k \cfrac{f_{Y;s}(\cdot)}{F_{Y;k}(\cdot)}F_{D;s}(\cdot)$.
Notice that $\sum_{s=1}^k \cfrac{f_{Y;s}(\cdot)}{F_{Y;k}(\cdot)} = 1$, since the denominator is the value of the cumulative distribution of the numerator at the end of the summation interval.
Lastly, the equation \eqref{eq_te_E} is obtained by rearranging the terms:
\begin{equation*}
\begin{aligned}
\psi_k^{\text{TE}} &= F_{Y;k}(a) [1-E_k(a,a)] - F_{Y;k}(a^*) [1-E_k(a^*,a^*)] \\[1ex]
&= F_{Y;k}(a) - F_{Y;k}(a)E_k(a,a) - F_{Y;k}(a^*) + F_{Y;k}(a^*)E_k(a^*,a^*) \\[1ex]
&= [F_{Y;k}(a) - F_{Y;k}(a^*)] - [F_{Y;k}(a)E_k(a,a) - F_{Y;k}(a^*)E_k(a^*,a^*)].
\end{aligned}
\end{equation*}

\section{4-way decomposition}\label{app_decomp}

Based on the relationship $Y_k^{a',a''}=Y_k^{a',\varnothing}[1-D_k^{a',a''}]$ (see Section 4.1. of the main text), the four estimands  \eqref{eq_yd} of the main text
have the following expressions:

$$
\begin{aligned}
\text{CDE}_k &= Y_k^{a,\varnothing} - Y_k^{a^*,\varnothing},\\[1ex]
\text{INT}_{\text{ref},k} &=
Y_k^{a,\varnothing}(1-D_k^{a,a^\ast}) - 
Y_k^{a^*,\varnothing}(1-D_k^{a^*,a^*})
- (Y_k^{a,\varnothing} - Y_k^{a^*,\varnothing})\\[1ex] 
& = -(Y_k^{a,\varnothing}D_k^{a,a^*} - Y_k^{a^*,\varnothing} D_k^{a^*,a^*}),\\[1ex]
\text{INT}_{\text{med},k} &= Y_k^{a,\varnothing}(1-D_k^{a,a}) - Y_k^{a^*,\varnothing}(1-D_k^{a^*,a}) \\
& \hspace{3em} - \bigl\{Y_k^{a,\varnothing}(1-D_k^{a,a^*}) - Y_k^{a^*,\varnothing}(1-D_k^{a^*,a^*})\bigr\}\\[1ex]
&= - (Y_k^{a,\varnothing}D_k^{a,a} - Y_k^{a^*,\varnothing}D_k^{a^*,a}) + (Y_k^{a,\varnothing}D_k^{a,a^*} - Y_k^{a^*\varnothing}D_k^{a^*,a^*}) \\[1ex]
\text{PIE}_k  &=  Y_k^{a^*,\varnothing}(1-D_k^{a^*,a})- Y_k^{a^*,\varnothing}(1-D_k^{a^*,a^*}) =  -Y_k^{a^*,\varnothing}(D_k^{a^*,a} - D_k^{a^*,a^*}).
\end{aligned}
$$

\section{Identification proofs}\label{appendix_identification}
Let $\bar Y_k^{a',\varnothing}=(Y_0^{a',\varnothing},\ldots,Y_k^{a',\varnothing})$ and $\bar D_k^{\varnothing,a''}=(D_0^{\varnothing,a''},\ldots,D_k^{\varnothing,a''})$ denote the histories of the controlled processes up to time $t_k$. Correspondingly, $\underline Y_k^{a',\varnothing} = (Y_k^{a',\varnothing},\dots,Y_K^{a',\varnothing})$ and $\underline D_k^{\varnothing,a'}=(D_k^{\varnothing,a'},\dots,D_K^{\varnothing,a'})$ are the future
values of the controlled processes from time $t_k$ onward. $C_0=D_0=Y_0=0$ by definition.

For all $k=0,\dots,K-1$, and $a',a'' \in \{a,a^*\}$, we make the following assumptions:

\begin{description}
\item[A1] $(\bar Y_K^{a',\varnothing},\bar D_K^{\varnothing, a'}) \ind A\cond W $
\item[A2] $Y_{k+1}^{a',\varnothing} \ind D_{k+1}^{\varnothing,a''}\cond Y_k^{a',\varnothing}=0,W$
\item[A3] $D_{k+1}^{\varnothing,a''} \ind Y_{k}^{a',\varnothing}\cond D_k^{\varnothing,a''}=0,W$
\item[S1] $(\underline Y_{k+1}^{a',\varnothing},\underline D_{k+1}^{\varnothing,a'}) \ind C_{k+1}\cond \bar C_k = D_k = Y_k = 0, A=a', W$
\item[C1] $D_k^{a'}=0 \implies Y_k^{a'} = Y_k^{a',\varnothing}, 
~k=0,\dots,K$
\item[C2] $Y_k^{a'}=0 \implies D_{k+1}^{a'}=D_{k+1}^{\varnothing,a'},
~k=0,\dots,K-1$
\item[C3] $A=a' \implies D_k = D_k^{a'} \text{ and } Y_k = Y_k^{a'}$
\item[P1] $P(Y_k=D_k=0\cond \bar C_k = Y_{k-1}=D_{k-1}=0,A=a',w) > 0$
\end{description}
Assumption A1 states that, conditionally on the baseline covariates $W$, the two controlled processes are jointly independent of treatment assignment $A$.
In experimental studies, this is ensured through random allocation of treatment.
The sequential exchangeability assumptions A2 and A3 require that, given the baseline covariates and survival from either process until time $t_k$, the next two possible events are conditionally independent.
Assumption S1 imposes non-informative censoring.
Assumption P1 is a positivity assumption requiring that under either treatment, it is possible  to survive up to the last time $t_K$ for $w\in W$.
Assumption C3 is the standard consistency assumption.
Notice that the assumptions C1--C3 together imply that
$$
\begin{aligned}
A=a', D_k=Y_k=0 \implies D_{k+1} = D_{k+1}^{\varnothing,a'}, \\
A=a',D_{k+1} = Y_{k}=0 \implies Y_{k+1}=Y_{k+1}^{a',\varnothing}.
\end{aligned}
$$
\vspace{1em}

We first prove that the cross-world estimand $P(Y_k^{a',a''}=1)$ is identified under the exchangeability assumptions A1--A3, non-informative censoring S1, consistency C1--C3 and positivity P1. 
Subsequently, we show that the controlled-effect estimand $P(Y_k^{a',\varnothing}=1)$ is identified under the same set of assumptions, excluding assumption A3.
We then proceed to show that the identifiable expressions coincide with those of the empirical decomposition \eqref{eq_decomp_empirical} given in Section 3 of the main text.

\subsubsection*{Identification of cross-world estimands}
The proof of the cross-world case begins by presenting the probability of event $Y_k^{a',a''} = 1$ as a sum of point probabilities up to time $t_k$ (step 1) and invoking the definition of the joint process to express everything in terms of controlled processes (step 2). We then factorise with respect to the baseline covariates $W$ (step 3) and further with respect to the histories of the controlled processes (step 4). Of note, survival up to any time point implies survival through all previous time points.
Steps 1--4 lead to the following expressions:

$$
\begin{aligned}
& P(Y_k^{a',a''}=1) \\
&^{1.}= \sum_{s=1}^kP(Y_s^{a',a''}=1, D_s^{a',a''}=Y_{s-1}^{a',a''}=\dots=D_1^{a',a''}=Y_1^{a',a''}=0) \\
&^{2.}= \sum_{s=1}^kP(Y_s^{a',\varnothing}=1, D_s^{\varnothing,a''}=Y_{s-1}^{a',\varnothing}=\dots=D_1^{\varnothing,a''}=Y_1^{a',\varnothing}=0)\\
&^{3.}= \sum_{w}\sum_{s=1}^{k}P(Y_s^{a',\varnothing}=1, D_s^{\varnothing,a''}=Y_{s-1}^{a',\varnothing}=\dots=D_1^{\varnothing,a''}=Y_1^{a',\varnothing}=0,w)\\
&^{4.}= \sum_{w}\sum_{s=1}^{k}P(Y_s^{a',\varnothing}=1\cond  D_s^{\varnothing,a''}= Y_{s-1}^{a',\varnothing} = 0,w)\\
& ~ \hspace{2cm}\times\prod_{j=1}^{s} P(D_j^{\varnothing,a''}=0\cond D_{j-1}^{\varnothing,a''}=Y_{j-1}^{a',\varnothing}=0,w) \\ 
& ~ \hspace{3cm}\times P(Y_{j-1}^{a',\varnothing}=0\cond Y_{j-2}^{a',\varnothing} = D_{j-1}^{\varnothing,a''}=0,w)P(w).
\end{aligned}
$$

Next, we employ assumptions A2 and A3 to change treatments levels under which the survival from the controlled processes are conditioned on (step 5). Assumption A1 along with positivity P1 then allows conditioning each term also on the observed treatment assignment (step 6).
\begin{equation}\label{eq_SWtelescope}
\begin{aligned}
& P(Y_k^{a',a''}=1) \\
&^{5.}= \sum_{w}\sum_{s=1}^{k}P(Y_s^{a',\varnothing}=1\cond  D_s^{\varnothing,a'}= Y_{s-1}^{a',\varnothing} = 0,w)\\
& ~ \hspace{2cm}\times \prod_{j=1}^{s} P(D_j^{\varnothing,a''}=0\cond D_{j-1}^{\varnothing,a''}=Y_{j-1}^{a'',\varnothing}=0,w) \\ 
& ~ \hspace{3cm} \times P(Y_{j-1}^{a',\varnothing}=0\cond Y_{j-2}^{a',\varnothing} = D_{j-1}^{\varnothing,a'}=0,w)P(w)\\
&^{6.}= \sum_{w}\sum_{s=1}^{k}P(Y_s^{a',\varnothing}=1\cond  D_s^{\varnothing,a'}= Y_{s-1}^{a',\varnothing} = 0,A=a',w)\\
& ~ \hspace{2cm} \times \prod_{j=1}^{s} P(D_j^{\varnothing,a''}=0\cond D_{j-1}^{\varnothing,a''}=Y_{j-1}^{a'',\varnothing}=0,A=a'',w) \\ 
& ~ \hspace{3cm} \times P(Y_{j-1}^{a',\varnothing}=0\cond Y_{j-2}^{a',\varnothing} = D_{j-1}^{\varnothing,a'}=0,A=a',w)P(w).
\end{aligned}
\end{equation}

To incorporate censoring, we include the trivial condition $D_0=Y_0=C_0=0$. We then invoke the assumptions P1, S1 and C1--C3 repeatedly to express each term using observable quantities.
Consider the probability $P(Y_k^{a',\varnothing}=y \cond  D_k^{\varnothing,a'}=Y_{k-1}^{a',\varnothing}=0,A=a',w)$, $k=1,\dots,K,$ and $y\in\{0,1\}$, and recall that $Y_{k-1}^{a',\varnothing}=0$ implies
$\bar Y_{k-1}^{a',\varnothing}=0$,
and $D_{k-1}^{\varnothing,a''}=0$ implies
$\bar D_{k-1}^{\varnothing,a''}=0$.
We write

$$
{\small
\begin{aligned}
P&(Y_k^{a',\varnothing}=y \cond  D_k^{\varnothing,a'}=Y_{k-1}^{a',\varnothing}=0,A=a',w)\\
&^{7.}=  \frac{P(Y_k^{a',\varnothing}=y, D_k^{\varnothing,a'} = Y_{k-1}^{a',\varnothing}=\dots=D_1^{\varnothing,a'} = Y_1^{a',\varnothing}=0\cond C_0 = D_0 = Y_0 = 0, A=a',w)}{P(D_k^{\varnothing,a'} = Y_{k-1}^{a',\varnothing}=\dots=D_1^{\varnothing,a'} = Y_1^{a',\varnothing}=0\cond C_0 = D_0 = Y_0 = 0, A=a',w)} \\
&^{8.}=  \frac{P(Y_k^{a',\varnothing}=y, D_k^{\varnothing,a'} = Y_{k-1}^{a',\varnothing}=\dots=D_1^{\varnothing,a'} = Y_1^{a',\varnothing}=0\cond \bar C_1 = D_0 = Y_0 = 0, A=a',w)}{P(D_k^{\varnothing,a'} = Y_{k-1}^{a',\varnothing}=\dots=D_1^{\varnothing,a'} = Y_1^{a',\varnothing}=0\cond \bar C_1 = D_0 = Y_0 = 0, A=a',w)} \\
&^{9.}=  \frac{P(Y_k^{a',\varnothing}=y, D_k^{\varnothing,a'} = Y_{k-1}^{a',\varnothing}=\dots=D_2^{\varnothing,a'} = Y_1^{a',\varnothing}=0\cond \bar C_1 = D_1 = Y_0 = 0, A=a',w)}{P(D_k^{\varnothing,a'} = Y_{k-1}^{a',\varnothing}=\dots=D_2^{\varnothing,a'} = Y_1^{a',\varnothing}=0\cond \bar C_1 = D_1 = Y_0 = 0, A=a',w)} \\
&^{10.}= \frac{P(Y_k^{a',\varnothing}=y, D_k^{\varnothing,a'} = Y_{k-1}^{a',\varnothing}=\dots=D_2^{\varnothing,a'} = Y_2^{a',\varnothing}=0\cond \bar C_1 = D_1 = Y_1 = 0, A=a',w)}{P(D_k^{\varnothing,a'} = Y_{k-1}^{a',\varnothing}=\dots=D_2^{\varnothing,a'} = Y_2^{a',\varnothing}=0\cond \bar C_1 = D_1 = Y_1 = 0, A=a',w)}
\end{aligned}
}
$$
Step 7 used the positivity assumption to ensure the denominator is not zero. 
Step 8 used S1 to include conditioning on $\bar C_1=0$.
Steps 9 and 10 used the consistency assumptions C1--C3 to first change the indicator $D_1^{\varnothing,a'}$ and then $Y_1^{a',\varnothing}$ to their observed counterparts.
Repeating these steps $k-1$ times we have
\begin{multline}\label{eq_cens}
P(Y_k^{a',\varnothing}=y \cond  D_k^{\varnothing,a'}=Y_{k-1}^{a',\varnothing}=0,A=a',w) \\
= P(Y_k=y\cond D_k = Y_k = \bar C_k = 0, A=a', w).
\end{multline}
A similar procedure may be straightforwardly applied to the probabilities concerning the competing ($D$) event.

Finally, applying \eqref{eq_cens} to all of the terms in \eqref{eq_SWtelescope} the cross-world estimand is expressed as
\begin{equation}\label{app_idres}
P(Y_k^{a',a''}=1) = \sum_{s,w} \lambda_{Y;s}(a',w)\prod_{j=1}^s\big[1-\lambda_{Y;j-1}(a',w)\big]\big[1-\lambda_{D;j}(a'',w)\big]P(w),
\end{equation}
where the discrete-time cause-specific hazards
\begin{equation*}
\begin{aligned}
\lambda_{Y;l}(a''') &=  P(Y_l=1\cond D_l=Y_{l-1}=\bar C_l = 0,A=a''')\hspace{4mm}\text{and}\\
\lambda_{D;l}(a''') &= P(D_l=1\cond Y_{l-1}=D_{l-1}=\bar C_l=0,A=a''')
\end{aligned}
\end{equation*}
can be estimated from observed data.

\subsubsection*{Identification of controlled estimands}
In the controlled estimand case we require assumption A2 only under the special case $a' = a''$ and drop assumption A3.
The proof proceeds similarly to that in the cross-world case.
We start by writing the probability of the event $Y_k^{a',\varnothing}=1$ as a sum of point probabilities and summing over all covariate strata $w$.
We then factorise the joint distribution into sequential conditional probabilities and invoke the assumption A2 to condition on survival from the controlled $D$ process under the same treatment.
This leads to the following:
\begin{equation}\label{eq_controlled_proof}
\begin{aligned}
& P(Y_k^{a',\varnothing}=1) \\
&= \sum_w\sum_{s=1}^{k}P(Y_s^{a',\varnothing}=1, Y_{s-1}^{a',\varnothing}=\dots Y_1^{a',\varnothing}=0, w) \\
&= \sum_w\sum_{s=1}^{k}P(Y_s^{a',\varnothing}=1\cond  Y_{s-1}^{a',\varnothing}=0,w) \\
& ~ \hspace{2cm} \prod_{j=1}^{s-1}P(Y_j^{a',\varnothing} = 0 \cond  Y_{j-1}^{a',\varnothing}=0, w)P(w) \\
&= \sum_w\sum_{s=1}^{k}P(Y_s^{a',\varnothing}=1\cond  D_s^{\varnothing,a'} = Y_{s-1}^{a',\varnothing}=0,w) \\
& ~ \hspace{2cm} \prod_{j=1}^{s-1}P(Y_j^{a',\varnothing} = 0 \cond D_j^{\varnothing,a'} = Y_{j-1}^{a',\varnothing}=0, w)P(w).
\end{aligned}
\end{equation}

For any $l = 1,\dots,K$ and $y\in\{0,1\}$ we have
\begin{equation}\label{eq_addA}
\begin{aligned}
&P(Y_l^{a',\varnothing}=y\cond D_l^{\varnothing,a'}=Y_{l-1}^{a',\varnothing}=0,w)\\[1ex]
&= \frac{P(Y_l^{a',\varnothing}=y,D_{l}^{\varnothing,a'} = Y_{l-1}^{a',\varnothing} = 0 \cond  w)}{P(D_{l}^{\varnothing,a'} = Y_{l-1}^{a',\varnothing} = 0 \cond  w)} \\[1ex]
&= \frac{P(Y_l^{a',\varnothing}=y,D_{l}^{\varnothing,a'} = Y_{l-1}^{a',\varnothing} = 0 \cond  A=a', w)}{P(D_{l}^{\varnothing,a'} = Y_{l-1}^{a',\varnothing} = 0 \cond  A=a', w)} \\[1ex]
&= P(Y_l^{a',\varnothing} = y\cond D_l^{\varnothing,a'}=Y_{l-1}^{a',\varnothing}=0,A=a',w),
\end{aligned}
\end{equation}
where the first equality requires the positivity assumption P1 and the second invokes the exchangeability A1.

Applying Eqs. \eqref{eq_addA} and \eqref{eq_cens} to \eqref{eq_controlled_proof} we obtain
\begin{equation}\label{eq_idres_controlled}
\begin{aligned}
P(Y_k^{a',\varnothing} = 1) &= \sum_w \sum_{s=1}^{k}\lambda_{Y;s}(a',w)\prod_{j=1}^{s-1}\big[1-\lambda_{Y;s}(a',w)\big]P(w),
\end{aligned}
\end{equation}
which concludes the proof.

\subsubsection*{Identification of the decomposition components}

Identification of the decomposition components follows trivially from the identification of the cross-world and controlled estimands.
To express the average four-way decomposition estimands in terms of observed data distribution, we replace the cross-world estimands $Y_k^{a',a''}$ and the controlled estimands $Y_k^{a',\varnothing}$ in Eq. \eqref{eq_decomposition} of the main text with their observable expectations \eqref{app_idres} and \eqref{eq_idres_controlled}.
This leads directly to the expressions id1--id4 in the main text. 

Furthermore, to see that the five expressions coincide with those in equations 
\eqref{eq_decomp_empirical} of the main text, note that
$$
\begin{aligned}
&\sum_{s=1}^k \lambda_{Y;s}(a',w) \prod_{j = 1}^s\big[ 1-\lambda_{Y;j-1}(a',w) \big]\big[1-\lambda_{D;j}(a'',w) \big] \\
&= \sum_{s=1}^k f_{Y;s}(a',w)\big[1-F_{D;s}(a'',w)] = F_{Y;k}(a',w) - \sum_{s=1}^k f_{Y;s}(a',w)F_{D;s}(a'',w),
\end{aligned}
$$
where $f_{Y;s}(\cdot)$ is the net point probability that the target event ($Y$) occurs at time $t_s$ and $F_{Y;l}(\cdot)$ and $F_{D;l}(\cdot)$ are the net risks of the two events up to time $t_l$.
Multiplying and dividing the second term above by $F_{Y;k}(a',w)$ gives
$$
\begin{aligned}
&F_{Y;k}(a',w) - \sum_{s=1}^k f_{Y;s}(a',w)F_{D;s}(a'',w) \\
& = F_{Y;k}(a',w) - F_{Y;k}(a',w)\sum_{s=1}^k\cfrac{f_{Y;s}(a',w)}{F_{Y:k}(a',w)}F_{D,s}(a'',w) \\
&= F_{Y;k}(a',w) - F_{Y;k}(a',w)E(a',a'',w),
\end{aligned}
$$
where 
$$
E_k(a',a'',w) = \sum_{s=1}^k \tilde f_{Y_k}(s\cond a',w){F_{D;s}(a'',w)} = \E[S \sim \tilde f(s\cond a',w,Y_k=1)]{F_{D;S}(a'', w)}
$$
and
$\tilde f(s\cond \cdot,Y_k=1) = f_{Y,s}(\cdot)/F_{Y;k}(\cdot)$.
From this, equations \eqref{eq_decomp_empirical} follow directly.

\section{Restricted mean survival time}\label{app_rmst}

The restricted mean survival time (RMST) is the average time spent event-free over a preset time horizon \cite{Royston2013}.  
With time horizon $t_k\in\lbrace t_0,\ldots,t_K\rbrace$, the RMST for a time-to-event variable $T$ is

$$
\E{R_k} = \E{\min\{ t_k, T\}} = t_k - \sum_{s=1}^k \Delta_s P(T < s),
$$
where $\Delta_s = t_s - t_{s-1}$.
The RMST corresponding to the event of interest $Y$ of a joint process under treatment levels $a'$ and $a''$ is thus
\begin{equation}\label{eq_rmstdef}
\E{R^{a',a''}_k} = t_k - \sum_{s=1}^k \Delta_s \E{Y^{a',a''}_s}.
\end{equation}
Furthermore, by equation \eqref{eq_rmstdef}, any causal effect estimand defined on the cumulative risk scale as differences of counterfactual risks, translates into the RMST scale.
For example:
\begin{equation*}\label{eq_rmst}
\mathbb{E}[R_k^{a',a''}] - \mathbb{E}[R_k^{a'',a''}] 
= -\sum_{s=1}^k\Delta_s(\mathbb{E}[Y_k^{a',a''}] - \mathbb{E}[Y_k^{a'',a''}]).
\end{equation*}
It follows that the total effect and its four-way decomposition can be written on the RMST scale as:
$$
\begin{aligned}
\text{RTE} &= - \sum_{s=1}^k \Delta_s \text{TE}_s,\\
\text{RCDE} &= - \sum_{s=1}^k \Delta_s\times \text{CDE}_s,\\
\text{RINT}_{\text{ref}} &= - \sum_{s=1}^k \Delta_s \times \text{INT}_{\text{ref},s},\\
\text{RINT}_{\text{med}} &= -\sum_{s=1}^k \Delta_s \times \text{INT}_{\text{med},s}, \\
\text{RPIE} &= -\sum_{s=1}^k \Delta_s \times \text{PIE}_s.
\end{aligned}
$$
An advantage of using the restricted mean survival time as a causal effect measure is that it captures the accumulated effects over the follow-up, summarising them into interpretable scalar values. In contrast, the cumulative risk scale does not account for the timing of events.

\section{Empirical expressions for NDE, NIE and TDE}\label{app_comb}

As introduced in section \ref{section_choosing}, the natural direct effect ($\text{NDE} = \text{CDE}+\text{INT}_\text{ref}$), natural indirect effect ($\text{NIE} = \text{INT}_\text{med}+\text{PIE}$) and total direct effect ($\text{TDE} = \text{INT}_\text{ref}+\text{INT}_\text{med}+\text{PIE}$) are obtained by summing components of the four-way decomposition.
A straightforward calculation using the results id2--id5 in the main text gives the empirical expressions for the average NDE, NIE and TDE as:
\begin{equation*}
\begin{aligned}
&\E{\text{NDE}_k} = \E{\text{CDE}_k} + \E{\text{INT}_{\text{ref},k}}\\ 
&= \sum_w \sum_{s=1}^k\Big(\lambda_{Y;s}(a,w) \prod_{j=1}^s\big[1-\lambda_{Y;j-1}(a,w)\big]\big[1 - \lambda_{D;j}(a^*,w) \big] \\
& \qquad - \lambda_{Y;s}(a^*,w)\prod_{j=1}^s\big[ 1 - \lambda_{Y;j-1}(a^*,w)\big] \big[ 1-\lambda_{D;j}(a^*,w) \big] \Big)P(w) \\[1ex]
&\E{\text{NIE}_k} = \E{\text{INT}_{\text{med},k}} + \E{\text{PIE}}_k \\
&= \sum_w \sum_{s=1}^k\Big(\lambda_{Y;s}(a,w) \prod_{j=1}^s\big[1-\lambda_{Y;j-1}(a,w)\big]\big[1 - \lambda_{D;j}(a,w) \big] \\
& \qquad - \lambda_{Y;s}(a,w)\prod_{j=1}^s\big[ 1 - \lambda_{Y;j-1}(a,w)\big] \big[ 1-\lambda_{D;j}(a^*,w) \big] \Big)P(w) \\[1ex]
&\E{\text{TDE}_k} = \E{\text{CDE}_k} + \E{\text{INT}_{\text{ref},k}} + \E{\text{INT}_{\text{med},k}} \\
&= \sum_w \sum_{s=1}^k\Big(\lambda_{Y;s}(a,w) \prod_{j=1}^s\big[1-\lambda_{Y;j-1}(a,w)\big]\big[1 - \lambda_{D;j}(a,w) \big] \\
& \qquad - \lambda_{Y;s}(a^*,w)\prod_{j=1}^s\big[ 1 - \lambda_{Y;j-1}(a^*,w)\big] \big[ 1-\lambda_{D;j}(a,w) \big] \Big)P(w).
\end{aligned}
\end{equation*}

\end{document}